\documentstyle[aps, psfig, manuscript]{revtex}

\begin{document}
\title{Quantum non-demolition measurements using Bose-Einstein
condensates}
\author{S. Choi and K. Burnett}
\address{Clarendon Laboratory, Department of Physics, University of
Oxford, Parks Road,\\
Oxford OX1 3PU, United Kingdom}
\maketitle

\begin{abstract}
We study possible scenarios for quantum non-demolition measurements using
Bose-Einstein condensates. We show that the interactions between condensate atoms makes it possible to measure the atom number with minimal back action on the system. This is an example of atomic nonlinear optics that has significant implications in the quantum state preparation of the Bose condensates.
\end{abstract}


\vspace{6mm}

\section{Introduction}

The concept of quantum non-demolition (QND) measurement was originally
introduced in the context of the detection of the gravitational waves\cite{Caves1} which produce very weak changes in the measuring instrument. 
In general, the quantum mechanical unpredictability disturbs the system being measured. This presents particular challenges in the context of ultra-sensitive measurements where the disturbances may mask the real effects to be measured. A sequence of {\it completely predictable} measurements on a quantum system, in the sense that the measurements do not disturb the system (``back-action evading''), is, however, possible for certain detector observables $\hat{A}(t)$, called QND observables, that satisfy the relation
$[\hat{A}(t),\hat{A}(t')]=0$, where $t$ and $t^{\prime }$ denote different times. In addition to its
relevance in ultra-sensitive measurements, a QND scheme provides a
way to prepare quantum mechanical states which may otherwise be difficult to
create, such as Fock states with a specific number of particles. In the
context of research in Bose-Einstein condensation (BEC), such quantum
state preparation procedure would be particularly valuable, for instance, in
engineering entangled state or Schr\"{o}dinger's cat states. Indeed, this is perhaps the most important motivation in searching for QND schemes with atoms. It is noted that a state preparation with BEC has recently been performed in the form of squeezed state creation in an optical lattice\cite{kasevich}. A successful QND scheme would create an important quantum state in the form of the Fock states of known number of atoms.

One of the original proposals for a quantum optical QND scheme
was that involving the Kerr medium\cite{walls2}, which changes its
refractive index as a function of the number of photons in the
``signal'' pump laser. We wish to apply the same strategy with the atoms
where the nonlinearity is intrinsic i.e. due to atomic interactions. The advent of experimental methods for producing BEC should, in principle, enable us to make progress in the matter-wave analogue of the optical QND experiments. 
In particular, the presence of the long range order in a BEC implies that the condensate phase can be accurately measured. 
The schemes to be proposed are based on experimental technologies available immediately: (1) the Bragg scattering of the condensates\cite{Kozuma}, and (2) creation of multicomponent, trapped Bose condensates in different hyperfine states\cite{JILA2,MIT}.

\section{QND using Bragg scattering}

Consider a QND scheme in which two light fields are used to produce equal parts of a trapped BEC which is Bragg scattered into distinct momenta $\hbar k_{p}$ and $-\hbar k_{p}$ in the center-of-mass (CM) frame\cite{Kozuma}.  The momentum $\hbar k_{p}$ may be any multiple of unit momentum kick $\hbar k_{R}$, $\hbar k_{p} = n\hbar k_{R}$, where $k_{R}$ is the wave number of the Raman laser. The relevant parts of the field operator at the end of the Bragg pulse may be expanded as a superposition of two counterpropagating modes of momenta $\pm \hbar k_{p}$ in this CM frame as
\begin{equation}
\hat{\Psi}(x)=\frac{1}{\sqrt{V}}(\hat{a}_{p}e^{ik_{p}x}+\hat{a}_{-p}e^{-ik_{p}x}),
\end{equation}
where $V$ is the volume of confinement, and $\hat{a}_{p}$,
$\hat{a}_{-p}$ denote the annihilation operators of the two counterpropagating modes.  
Evolution of this Bragg scattered state under the standard many-body boson Hamiltonian provides interatomic collisions which are necessary for the QND measurement.

The effective Hamiltonian for this system is then given by
\begin{equation}
\hat{{\cal H}}_{\rm Eff}=\frac{\hbar ^{2}k_{p}^{2}}{2M}(\hat{a}_{p}^{\dagger
}\hat{a}_{p}+\hat{a}_{-p}^{\dagger}\hat{a}_{-p})+ U_{+p} \hat{a}_{p}^{\dagger }\hat{a}_{p}^{\dagger }\hat{a}_{p}\hat{a}_{p} + U_{-p}\hat{a}_{-p}^{\dagger }\hat{a}_{-p}^{\dagger
}\hat{a}_{-p}\hat{a}_{-p}+ U_{\pm p} \hat{a}_{p}^{\dagger }\hat{a}_{p}\hat{a}_{-p}^{\dagger }\hat{a}_{-p}.
\label{hamiltonian}
\end{equation}
The Hamiltonian is similar to the well-known quantum optical version due to a Kerr medium, except for the self-interaction terms i.e. the 3rd and 4th terms. We have deliberately distinguished the self-collision and the interaction between the $+p$ and $-p$ modes by using $U_{+p}$, $U_{-p}$, and $U_{\pm p}$. We shall discuss this in more detail below. The first four terms represent two self-interacting condensates moving towards each other with momenta $-\hbar k_{p}$ and $\hbar k_{p}$ respectively, while the last term provides collisional coupling between the Boson number operators $\hat{a}_{p}^{\dagger }\hat{a}_{p}$ and $\hat{a}_{-p}^{\dagger }\hat{a}_{-p}$, mediated by the interatomic collisions. We shall refer to this last term as the interaction Hamiltonian, $\hat{{\cal H}}_{I}$.  We shall assume that any quasiparticle excitations created by the collision are minimal by assuming that we are in a slow moving regime ($k_{p}  \rightarrow 0$) and that the energies deposited by the collisions (as determined by the kinetic energy corresponding to the relative momenta, $2\hbar k_{p}$) do not match the energy of collective excitations of a confined condensate.

The collisional interactions are more accurately described by a representation of the two body $T$-matrix;  $U_{+p}$, $U_{-p}$ and $U_{\pm p}$ should therefore be calculated accordingly. In particular, in modelling  a specific experiment one has to treat the expansion of the condensate as well as the collisional loss of atoms ``knocked out'' by other passing atoms in order to assess the quality of QND schemes. A ``spherical shell'' of atoms is generated when two distinct momentum wave packets collide, due to elastic scattering into the many momentum-conserving channels; a significant portion of the atoms may consequently be lost from the characteristic mixing process. It should be possible to perform the whole measurement within a trap so that elastic losses are minimised. Recent calculations indicate that the fraction of atoms retained in the original wave packets approaches 90\% if the asymmetry of the trapping potentials is increased to an aspect ratio ($\omega_{z}/\omega_{x} = \omega_{z}/\omega_{y}$) of order 0.35\cite{Band}; we assume quasi 1-dimensional systems such as the cigar-shaped condensates in this work.   Using the result of Ref. \cite{Band}, we write $U_{+p} = U_{-p} = \frac{U_{0}}{2}$, where $U_{0} = \frac{4 \pi \hbar^2 a}{m}$ and $U_{\pm p} = 2U_{0}(1+i\gamma(k_{p}))$.

The imaginary part of $U_{\pm p}$ generates elastic losses proportional to the relative momentum of the wave packets for higher momenta. In this limit $\gamma(k) = 2k_{p}a$ where $a$ is the interatomic scattering length; this implies that a slower probe moving through the target is needed to minimise the elastic collisional losses.  The elastic collisional loss is further reduced by superfluidity effects, and energies in the phonon regime ($k_{p} \rightarrow 0$) could be used, as in this limit one has $\gamma(k) \propto k^{4}$.  In addition, with an initial estimate of the number, the losses can be calculated with high accuracy and we shall therefore assume the calculation is carried out in an iterative process where the number of atoms measured is fed back into the calculation.  We note that the interaction strength also changes in reality as a function of position so that the coupling in general depends on the shape of the overlapping beams as well as the composition of the counterpropagating wavepackets. It is reasonable to consider ideal cases of uniform and real valued effective interaction in this work, with $U_{\pm p} \equiv U_{+p}$ by virtue of $k_{p} \rightarrow 0$ and the factors discussed above.

In the absence of collisional loss, $\hat{a}_{p}^{\dagger }\hat{a}_{p}$ and $\hat{a}_{-p}^{\dagger
}\hat{a}_{-p}$ are both constants of the motion for this system, and hence any function of $\hat{a}_{\pm p}^{\dagger }\hat{a}_{\pm p}$ may be chosen to be the QND observable. We choose to label $\hat{{\cal O}}_{S}=\hat{a}_{p}^{\dagger }\hat{a}_{p}$ as the QND ``signal'' variable, and identify the probe variable to be the standard Hermitian phase quadrature operator, $\hat{{\cal O}}_{P}=i\left(
\hat{a}_{-p}-\hat{a}_{-p}^{\dagger }\right) $ where $\hat{a}_{-p}$ is such that $\hat{a}_{-p}\equiv \sqrt{\hat{a}_{-p}^{\dagger }\hat{a}_{-p}+1}\exp (i\hat{\varphi}_{-p})$, with $\hat{\varphi}_{-p}$ denoting the phase operator for $\hat{a}_{-p}$. Further necessary conditions such as $[\hat{{\cal O}}_{P},\hat{{\cal H}}_{I}]\neq 0$ and $[\hat{{\cal O}}_{S},\hat{{\cal H}}_{I}]=0$ are clearly satisfied.  $\hat{{\cal H}}_{I}$ is independent of the phase operator for the signal; this ensures the motion of $\hat{{\cal O}}_{S}$ do not become unpredictable due to the uncertainty imposed on the conjugate variable of $\hat{{\cal O}}_{S}$.

The Heisenberg equations of motion for the Boson annihilation operators $\hat{a}_{p}$ and $\hat{a}_{-p}$ are given by the coupled nonlinear operator equations
\begin{eqnarray}
i\frac{d\hat{a}_{p}}{dt} &=&\omega
_{p}\hat{a}_{p}+\tilde{U}_{0}\hat{a}%
_{p}^{\dagger }\hat{a}_{p}\hat{a}_{p}+2\tilde{U}_{0}\hat{a}%
_{-p}^{\dagger }\hat{a}_{-p}\hat{a}_{p}  \label{eqofmotion1} \\
i\frac{d\hat{a}_{-p}}{dt} &=&\omega
_{p}\hat{a}_{-p}+\tilde{U}_{0}%
\hat{a}_{-p}^{\dagger }\hat{a}_{-p}\hat{a}_{-p}+2\tilde{U}_{0}%
\hat{a}_{p}^{\dagger }\hat{a}_{p}\hat{a}_{-p}, 
\label{eqofmotion2}
\end{eqnarray}
where $\omega _{p}=\frac{\hbar k_{p}^{2}}{2M}$ and
$\hbar \tilde{U}_{0}=U_{0}/V$. In general, exact solutions to nonlinear
operator equations cannot be found. However, we have $\hat{a}_{\pm p}^{\dagger}\hat{a}_{\pm p}\hat{a}_{\mp p}=\hat{N}_{\pm p}\hat{a}_{\mp p}(t)$, where $\hat{N}_{\pm p}=\hat{a}_{\pm p}^{\dagger }\hat{a}_{\pm p}$ is a constant in time which enables us to transform to an interaction picture
$\hat{a}_{-p}\rightarrow \hat{a}_{-p}e^{-i(\omega _{p}+\tilde{U}_{0}\hat{N}_{-p})t}$ to write Eq.~(\ref{eqofmotion2}) as 
\begin{equation}
i\frac{d\hat{a}_{-p}}{dt}=2C_{p}\hat{a}_{-p},
\end{equation}
where $C_{p} = \tilde{U}_{0}\hat{a}_{p}^{\dagger
}\hat{a}_{p}$ is again a constant of the motion.
The solution is simply $\hat{a}_{-p}=\exp (-i2C_{p}t)
\hat{a}_{-p}(0)$, where $t$ is the duration of the collisional interaction between the ``signal'' and the ``probe'' condensates. A measurement of the probe phase therefore provides a QND information on the signal atom number for a given atomic species. For $^{87}{\rm Rb}$ samples $\langle 2C_{p} t\rangle \sim 10$ mrad, with $N_{p}\approx 5\times 10^{5}$ and $t\sim 1{\rm ms}$. One obtains $\frac{\Delta \phi }{\Delta N}\sim 2\times 10^{-2}t$ mrad per atom. The increase in sensitivity achievable by increasing $t$ may be compromised with the extra collisional losses in the system over time. This can be partially, but not fully, eliminated by modelling the loss process.

As a comparison, we note that a related way of performing a QND measurement would involve scattering single atoms off a Bose condensate, as a way to probe the cloud without disturbing it. We note that in the quantum mechanical framework, a particle of momentum $k$ incident on a potential $V(x)$ is analogous to light wave propagation in a medium with some refractive index\cite{QMI}. In a way reminiscent of a pumped Kerr medium, BEC then forms effectively a potential barrier as a function of the number of atoms within the condensate. We assume the condensate-induced potential is repulsive, although there are of course attractive cases.  If one assumes that the energy of the incident atoms is, say, twice the
peak energy of the condensate, the sensitivity of the phase shift, $\frac{\Delta \phi }{\Delta N}$, is calculated to be of the order of 10 mrad per atom for a typical atomic condensate with $N=5\times 10^{5}$\cite{footnote}.  A typical QND scheme may be visualised essentially as a Mach-Zehnder interferometry where the ``phase shifter'' in one arm is the target (or ``signal'') condensate, and the relative phase shift of the probe is measured with respect to a reference condensate split off at the first beam splitter. The phase of any given wave function is the sum of its spatial and temporal phases and in this particular example the pertinent phase shift which encodes the number is spatial.

We have also studied the dynamics of two colliding condensates have been simulated using a 1-dimensional Gross-Pitaevskii Equation (GPE). We have studied the evolution of two condensates within a harmonic trap in which one of the condensates is displaced from the center while the other is kept at the origin. This initial condition gives rise to a collisional interaction between the two in which the displaced condensate passes through the stationary one as a part of its simple harmonic motion. As the two condensates overlap at time $t = \frac{\pi}{2\omega}$ they form an interference pattern. At time $t = \frac{\pi}{\omega}$ the condensates again evolve into two distinct Gaussian-like profiles, although the original shapes are not restored. The dynamics of a reference condensate was calculated by carrying out an identical simulation but without the stationary condensate.  The phase shift at time $t = \frac{\pi}{\omega}$ was obtained numerically. It is noted that at $t = \frac{\pi}{\omega}$ the spatial derivative of the phase profile for both the reference and the colliding condensate is zero as the condensate momentarily becomes stationary at half period.  The simulation was repeated for a number of different values of the nonlinearity $C = NU_{0}$. A linear dependence of the phase shift is observed in accordance with the analysis above (Fig. 1).

A possible experimental scheme would involve the techniques of Mach-Zehnder Bragg interferometry using a sequence of local $\frac{\pi}{2}$-$\pi$-$\frac{\pi}{2}$ Bragg pulses\cite{Torii}. A similar experiment has also been performed at NIST where the trap was kept on. For a QND experiment a $\frac{\pi}{2}$-pulse in the place of a $\pi$-pulse may be applied on one arm of the interferometer. This would, in effect, generate a signal condensate out of the probe in one arm, and the number of atoms that has been ``pushed out'' may be ascertained from the resulting changes in the interference pattern.
The total time it takes to measure the probe, $t_{probe}$ should be shorter
than the time it takes for the condensate to evolve via self-interaction: $t_{probe}\ll m/N4\pi \hbar a\sim 0.1{\rm s}$ for $^{87}{\rm Rb}$.
From the reported experiment\cite{Torii}, the overall temporal and spatial scale of the experiment would be of the order $\tau \sim 50\mu {\rm s}$ (much shorted than the self evolution time) and $L\sim 1{\rm mm}$.

\section{QND schemes in multicomponent BEC}

A system of condensates in multiple hyperfine states is another candidate for QND measurement, and allows a potentially more flexible experimental geometry. Two overlapping condensates of Rubidium atoms in $F=1,m_{F}=-1$ and $F=2,m_{F}=2$ states have already been produced by the JILA group\cite{JILA2}, which allows coherent transfer of atoms from one state to another using appropriate sequence of radio-frequency (RF) and microwave transitions. On the other hand, the MIT group created three co-existing condensates in the ground state of sodium $F=1,m_{F}=-1,0,1$ by the use of an optical dipole trap\cite{MIT}. In order to take into account the condensates in different hyperfine levels, the trapping potential and the interatomic potentials need to be written in a more general form:  $V_{trap}=\sum_{i}V({\bf r}){\cal P}_{i}$ where ${\cal P}_{i}=|i\rangle \langle i|$ is the projector which projects a state into hyperfine level $i$, and for the interatomic interaction  $\delta ({\bf r}_{1}-{\bf r}_{2})\sum_{j}U_{j}{\cal P}_{j}$ where $U_{j}=4\pi a_{j}\hbar ^{2}/m$. The Hamiltonian for the multi-component system is then 
\begin{equation}
\hat{{\cal H}}=\hat{{\cal H}}_{A}+\hat{{\cal H}}_{B}+\hat{{\cal H}}_{int}, 
\end{equation}
where $\hat{{\cal H}}_{A,B}$ are the usual many-body Hamiltonians but with the field corresponding to atoms in magnetic levels $A,B$. The interaction Hamiltonian may be written in terms of mode operators $\hat{a}$ and $\hat{b}$ in the two mode (zero temperature) approximation: $\hat{{\cal H}}_{int}={\cal W}_{0}\hat{a}^{\dagger }\hat{a}\hat{b}^{\dagger }\hat{b}$. The atomic states are assumed to be  stable in the sense that they do not change their internal states on collisions.  The QND coupling term between the modes $\hat{a}^{\dagger }\hat{a}$ and $\hat{b}^{\dagger }\hat{b}$ then gives rise to
identical mathematical features as above. 
The Hamiltonian for the triplet $^{23}$Na system can be shown to couple atoms in all three states so that terms proportional to $\hat{a}_{-1}^{\dagger}\hat{a}_{-1}(\hat{a}_{1}^{\dagger }\hat{a}_{1}+\hat{a}_{0}^{\dagger }\hat{a}_{0})$,
$\hat{a}_{0}^{\dagger }\hat{a}_{0}(\hat{a}_{-1}^{\dagger }\hat{a}_{-1}+\hat{a}_{1}^{\dagger}\hat{a}_{1})$, and $\hat{a}_{1}^{\dagger
}\hat{a}_{1}(\hat{a}_{-1}^{\dagger }\hat{a}_{-1}+\hat{a}_{0}^{\dagger }\hat{a}_{0})$ appear in the Hamiltonian. Three possible QND signal/probe combinations may be exploited in this case, providing extra flexibility in the experiments. The detection of the probe phase requires separation of the two modes, which may be achieved by a Stern-Gerlach spin separation (or an RF output coupling). After the separation the phase may be measured e.g. by interfering with a previously prepared reference condensate.

We draw the attention to the fact that there are a couple of possible variations to the multicomponent QND scheme. One is the case where condensates in different hyperfine states are prepared in separate
traps $i$, $i=1,2$, which are then physically overlapped to produce
collisional interactions between the condensates. This type of procedure has previously been proposed in the context of quantum computation for individual atoms\cite{footnote2,Jaksch}. It is a more practical proposition for atoms held in optical traps, whose position can be manipulated with relative ease. The Hamiltonian is of the form 
\begin{equation}
\hat{{\cal H}} = \sum_{i=1}^{2} \hat{{\cal H}}_{A}^{i} + \sum_{j=1}^{2}\hat{{\cal H}}_{B}^{j} + \sum_{i,j=1}^{2}U_{i,j}\int d {\bf r}\hat{\Psi}_{A}^{i\dagger}({\bf r},t)\hat{\Psi}_{B}^{j\dagger}({\bf
r},t)\hat{\Psi}_{B}^{j}({\bf r},t)\hat{\Psi}_{A}^{i}({\bf r},t),
\end{equation}
where $\hat{{\cal H}}_{A(B)}^{i(j)}$ is the Hamiltonian for the component
$A(B)$ in the trap with trajectory ${\bf r}^{i(j)}(t)$, given by the single
particle Hamiltonians plus the nonlinear interactions. The probe and the signal
variables are identified by the different trap/hyperfine state combinations $\alpha,\beta,A,B$. Following collision-induced phase shift, the measurement of the phase may be carried out by either turning off the trap potential that contains the probe condensate or alternatively, by first displacing the trap with the probe. This is schematically illustrated in Fig. 2. Another possibility is where the traps of different trajectories contain two-species condensates in each. Phase changes in the atomic wave function may be induced such that two atoms in different hyperfine states $A$ and $B$ from traps with trajectories ${\bf r}^{\alpha}(t)$ and ${\bf r}^{\beta}(t)$ undergo a transformation $\psi^{\alpha}_{A}\psi^{\beta}_{B} \rightarrow e^{i\phi^{AB}}\psi^{\alpha}_{A}\psi^{\beta}_{B}$ where $\phi^{AB}$ is the collisional phase shift\cite{Jaksch}. For the case where the collisional phase shift $\phi^{AB}$ is $\pi$, one may write, using the Schwinger
angular momentum operators $\hat{J}_{x}^{i}$, $\hat{J}_{y}^{i}$,
the total Hamiltonian as
\begin{equation}
\hat{{\cal H}} = \Omega_{\alpha} \hat{J}_{z}^{\alpha} + \Omega_{\beta}
\hat{J%
}_{z}^{\beta} + R_{\alpha}\left( \hat{J}_{z}^{\alpha }\right) ^{2} +
R_{\beta}\left( \hat{J}_{z}^{\beta}\right) ^{2}+ R_{\alpha\beta}\hat{J}%
_{z}^{\alpha }\hat{J}_{z}^{\beta},
\end{equation}
where $\Omega, R_{\alpha}, R_{\beta}, R_{\alpha\beta}$ are constants. In this case, both $\hat{J}_{z}^{\alpha}$ and $\hat{J}_{z}^{\beta}$, the particle number difference of the two species in traps $\alpha$, $\beta$,  are constants of the motion.  The equations of motion for the angular momentum raising
operator $\hat{J}_{+}^{i} =\hat{a}_{i}^{\dagger}\hat{b}_{i}$ in the trap of trajectory $i$ are given by
\begin{eqnarray}
i\frac{d\hat{J}_{+}^{\alpha }}{dt} & = & \Omega_{\alpha} \hat{J}%
_{+}^{\alpha} + 2R_{\alpha}\hat{J}_{z}^{\alpha }\hat{J}_{+}^{\alpha }+
R_{\beta\alpha} \hat{J}_{z}^{\beta }\hat{J}_{+}^{\alpha } \\
i\frac{d\hat{J}_{+}^{\beta }}{dt} & = & \Omega_{\beta}\hat{J}^{\beta
}_{+} + 2 R_{\beta}\hat{J}_{z}^{\beta }\hat{J}_{+}^{\beta }+
R_{\alpha\beta}\hat{J}_{z}^{\alpha }\hat{J}_{+}^{\beta }.
\end{eqnarray}
The particle number difference $\hat{J}_{z}^{\alpha}$ in trap $\alpha$ may therefore be deduced from measuring $\hat{J}^{\beta}_{+}$, the phase difference between components $a_{i}$ and $b_{i}$ in trap $\beta$. $\hat{J}_{+}$ may be measured by passing the two condensates through an atom beam splitter and measuring the intensity at one of the output ports\cite{Kim}. All that is required is a way to couple out from the trap a superposition of the two components, $\hat{a}_{i} + e^{i\phi_{\rm rel}}\hat{b}_{i}$, where $\phi_{\rm rel}$ is the relative phase between the two components. To this end, one may first separate $\hat{a}_{i}$ and $\hat{b}_{i}$ from the ``probe'' trap into two separate traps (this is possible because they are in different hyperfine states) and interfere them normally by opening the traps while ``eliminating'' any information about from which trap an atom came from. This produces a spatial interference pattern which is exactly analogous to the case of having a beam splitter and an atom detector where the intensity at one of the output port oscillates in time (rather than space).

\section{Discussion}  
We have discussed several scenarios which enable QND measurement using the experimentally produced BEC in alkali atoms. The schemes proposed are already within current experimental capabilities, and should be achievable in the nearest future. They represent a new addition to the currently growing repertoire of atomic quantum optical experiments based on BEC. A recent experimental development that requires attention is the achievement of BEC in metastable He\cite{Aspect}; its potential sensitivity at the single atom level would be suitable for checking QND experiments measuring exact number of atoms.  A possible future application of the QND scheme is in quantum computation with BEC; it is essential to know the exact number of atoms in the condensate in order that the computational operations can be well-defined. Another example is in constructing atomic clocks with BEC; information about the exact number of atoms is essential in estimating the errors in the frequency. The QND schemes proposed are important in that they are all-atom based schemes which utilises the inherent collisions directly, rather than using light which may potentially heat or excite the ultracold atoms into different quantum states.  SC wishes to acknowledge support from the Royal Commission for the Exhibition of 1851 and KB the UK EPSRC.

\begin{figure}[t]
\centerline{\psfig{height=5cm,file=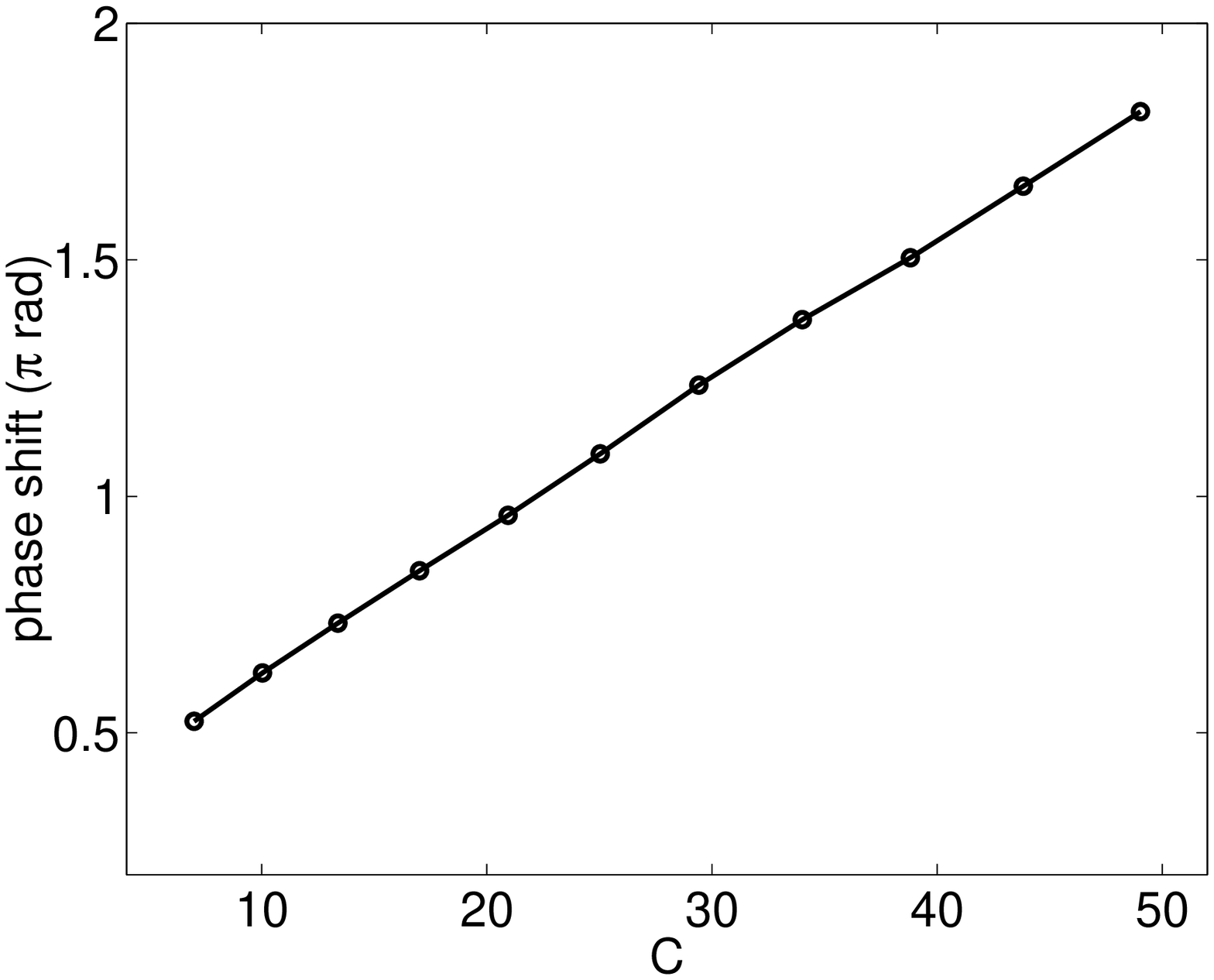}}
\caption{{\protect\footnotesize Phase shift gained by the displaced condensate after colliding with a stationary condensate at the origin. This is given as a function of the nonlinearity, $C$ which is directly proportional to the number of atoms in the condensate. The circles represent actual phase shift calculated using a GPE simulation.}}
\end{figure}

\begin{figure}[t]
\centerline{\psfig{height=5cm,file=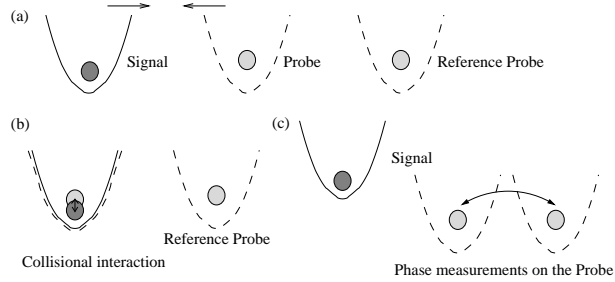}}
\caption{{\protect\footnotesize Diagram of a possible experimental
scheme.
(a) Two independent traps (indicated by the solid and dash-dot lines)
containing condensates in different hyperfine states are prepared and
are
brought towards each other. The probe and reference probe are made by splitting a single condensate. (b) The traps are spatially overlapped for
the
two condensates to interact. (c) One of the traps is then displaced and
appropriate measurements are made. }}
\end{figure}


\begin{references}
\bibitem{footnote}  This is, in fact, the current limit in best
precision phase measurement one may achieve with an atom interfermeter.
\bibitem{footnote2} The method of overlapping condensates confined in traps may be viewed as a way of mimicking the wave guides in optics, since atoms do not possess intrinsic translational motion.
\end{references}
\end{document}